\documentclass[NATO,numreferences]{crckbked}
\usepackage{epsfig}

\begin{raggedright}
\begin{opening}
\title{Nanostructured superconductor/\protect\\
       ferromagnet bilayers}

\author{M. Lange}
\author{M. J. Van Bael}
\author{L. Van Look}
\author{S. Raedts}
\author{V. V. Moshchalkov}
\author{Y. Bruynseraede}
 \institute{Laboratorium voor Vaste-Stoffysica an Magnetisme\\
            Katholieke Universiteit Leuven\\
            Celestijnenlaan 200D\\
            B-3001 Leuven
            Belgium}
\end{opening}
\end{raggedright}

\begin{document}

\section{Introduction}
Magnetic field can penetrate type-II superconductors in the form
of vortices. Each vortex carries a magnetic flux that is an
integer multiple of the flux quantum $\phi_{0}$. The pinning
properties of the vortices determine the magnitude of the critical
current density ($j_{c}$) and the magnetisation ($M$) of the
superconductor. Advances in nanolithography have allowed the
fabrication of superconducting thin films with artificial pinning
arrays like antidot lattices \cite{VVM,baert} or lattices of
magnetic dots \cite{martin,morgan,margriet,vanbaelPRB,vanbaelPRL}.
These pinning centres give rise to a huge enhancement of $j_{c}$
and $M$ and can be used to stabilize new vortex phases like
multiquanta and composite vortex lattices \cite{baert}. Pronounced
commensurability effects between the vortex lattice and the array
of pinning sites can be observed as peaks or cusps in $j_{c}(H)$
and $M(H)$ at specific values of the perpendicularly applied
magnetic field $H$.\\ We report on two different types of magnetic
pinning centres with out-of-plane magnetisation. In the first type
of sample, the Pb film is deposited on a square array of Co/Pt
multilayer {\em dots}, the second system consists of a Pb film
that is grown on a Co/Pt multilayer containing a regular array of
{\em antidots}. In both systems, we will investigate how the
direction of the magnetic moments in the Co/Pt multilayer
influences the flux pinning in the superconducting film. These
studies enable us to elucidate the pinning potential that the
magnetic nanostructures impose in the superconducting film.

\section{Sample preparation and characterization}
\subsection{Preparation}
\begin{figure}[tbp]
\centerline{\epsfig{file=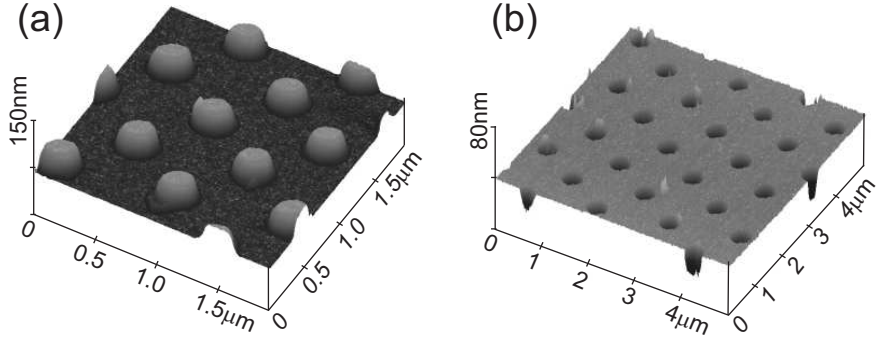}}
\nopagebreak \caption{AFM micrographs of a square array of
magnetic Co/Pt dots (a) and Co/Pt antidots (b). The dots
(Pt(6.4~nm)/[Co(0.5~nm)/Pt(1.6~nm)]$_{10}$) have a diameter of
0.26~$\mu$m, the lattice period amounts to 0.6~$\mu$m. The antidot
array (period 1~$\mu$m) is defined by square holes with a side
length of 0.37~$\mu$m in a
Pt(2.8~nm)/[Co(0.4~nm)/Pt(1.0~nm)]$_{10}$ multilayer.} \label{AFM}
\end{figure}

All samples were prepared on Si substrates with an amorphous
SiO$_{2}$ top layer. For preparation of the Co/Pt dots,
electron-beam lithography is used to define an array of holes in a
resist layer on the substrate. For fabrication of the magnetic
antidots, the resist on the substrate is predefined as an array of
dots. A Co/Pt multilayer is then evaporated in the resist mask
with a deposition rate of 0.01 nm/s for both Co and Pt at a
working pressure of $10^{-8}$~Torr. Finally the resist is removed
in a lift-off procedure, leaving an array of dots or antidots on
the substrate. Figure~\ref{AFM} shows two atomic force microscopy
(AFM) images of an array of Co/Pt dots and Co/Pt antidots.
Two Co/Pt dot samples were studied consisting of
[Co(0.5~nm)/Pt(1.6~nm)]$_{10}$ on a 6.4~nm Pt base layer and
[Co(0.4~nm)/Pt(0.9~nm)]$_{10}$ on a 2.5~nm Pt base layer. For both
dot arrays, the square lattice period is 0.6~$\mu$m and the dots
have the shape of a disk with 0.26~$\mu$m diameter.\\ The antidot
sample was made from a Co/Pt multilayer consisting of a 2.8~nm Pt
base layer and [Co(0.4~nm)/Pt(1.0~nm)]$_{10}$. The antidots have
square shape with rounded corners, their side length amounts to
0.37~$\mu$m, and they are arranged in a square array with period
1~$\mu$m.\\ After the magnetic properties of the samples were
characterized, they were covered with a Ge/Pb/Ge trilayer by
electron-beam evaporation at a working pressure of $10^{-8}$~Torr.
In order to prevent the direct influence of the proximity effects
between Pb and Co/Pt, a 10~nm insulating amorphous Ge layer is
deposited first with a growth rate of 0.2~nm/s, then the 50~nm Pb
film is evaporated at a substrate temperature of 77~K with a
growth rate of 1.0~nm/s and finally, the sample is covered with a
30~nm Ge layer for protection against oxidation. AFM images reveal
that smooth Pb layers are obtained, which completely cover the
dots and antidots. The critical temperature of the superconducting
Pb films is $T_{c}=7.20$~K.

\subsection{Magnetic characterization}
\begin{figure}[tbp]
\centerline{\epsfig{file=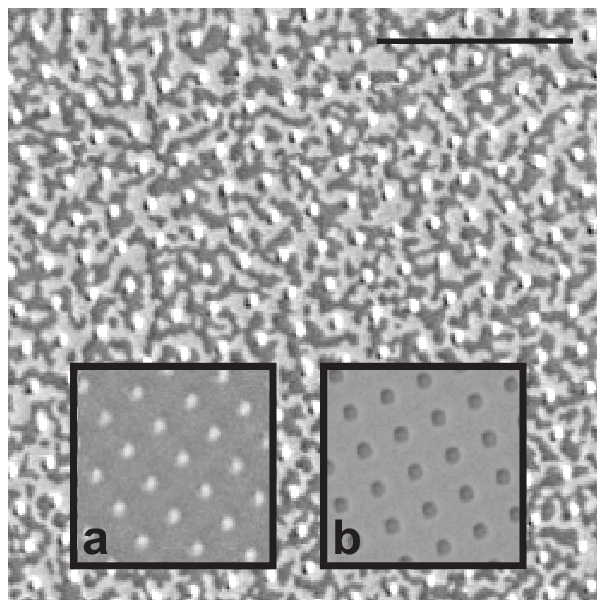}}
\caption{MFM image (15~$\mu$m $\times$ 15~$\mu$m) in $H=0$ of the
square array of Co/Pt antidots after out-of-plane demagnetisation;
the insets labeled by a and b show the remanent states after
magnetisation perpendicular to the film plane in a +10~kOe and
-10~kOe field, respectively. The black bar at the top of the image
corresponds to a length of 5~$\mu$m.} \label{mfmantidots}
\end{figure}

The easy axis of magnetisation in Co/Pt multilayers prepared with
correct film thicknesses lies perpendicular to the sample surface
\cite{zeper}. The out-of-plane anisotropy in our samples is
confirmed by magnetisation measurements using the magneto-optical
Kerr effect (MOKE) before deposition of the Ge/Pb/Ge trilayer.
\begin{figure}[tbp]
\centerline{\epsfig{file=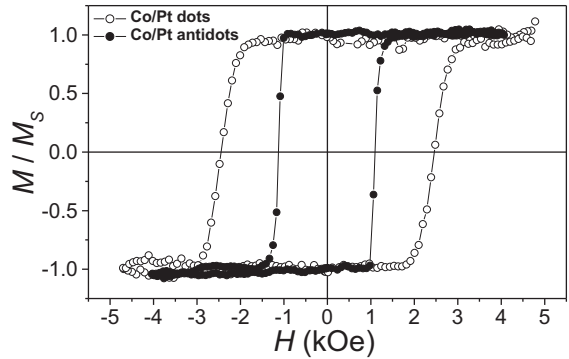}}
\caption{MOKE hysteresis loops of the
Pt(2.5~nm)/[Co(0.4~nm)/Pt(0.9~nm)]$_{10}$ dot array and the array
of antidots in a Pt(2.8~nm)/[Co(0.4~nm)/Pt(1.0~nm)]$_{10}$
multilayer (Co/Pt antidots) measured at room temperature and with
$H$ applied perpendicular to the sample surface.} \label{moke}
\end{figure}
Figure \ref{moke} shows MOKE hysteresis loops of the
Pt(2.5~nm)/[Co(0.4~nm)/Pt(0.9~nm)]$_{10}$ dot array and the Co/Pt
antidot array in a perpendicularly applied field $H$. A 100 \%
remanence is observed for both samples with coercive fields of
$H_{c}=1.1$~kOe for the antidots and $H_{c}=2.5$~kOe for the dots.
This difference in $H_{c}$ can be explained by the larger
demagnetisation factors of the Co/Pt antidots compared to the
dots. \\ The domain structure of the samples is investigated by
magnetic force microscopy (MFM).
\begin{figure}[tbp]
\centerline{\epsfig{file=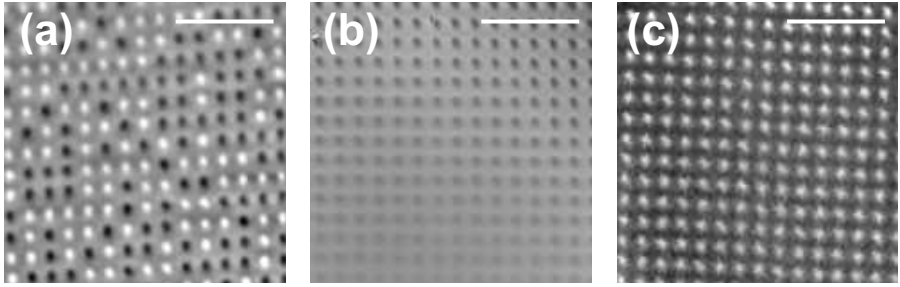}}
\caption{MFM images in zero field of a square array (period
0.6~$\mu$m) of Pt(2.5~nm)/[Co(0.4~nm)/Pt(0.9~nm)]$_{10}$ dots (a)
after out-of-plane demagnetisation; (b) and (c) in the remanent
state after magnetisation perpendicular to the film plane in a
-10~kOe and +10~kOe field, respectively. The white bar in all
images corresponds to a length of 3~$\mu$m.} \label{mfmdots}
\end{figure}
The MFM images in figure~\ref{mfmdots} show the dots in three
different remanent states, after demagnetisation in a
perpendicular field oscillating around zero with decreasing
amplitude (a), after magnetisation in a large negative field of
$H=-10$~kOe (b), and after magnetisation in a large positive field
of $H=+10$~kOe (c). After demagnetisation, all dots produce a
uniform dark or bright MFM signal, indicating a single domain
state with the magnetic moments $m$ either pointing up ($m>0$)
producing a bright signal, or pointing down ($m<0$), giving rise
to a dark signal. The average magnetic moment $\langle m \rangle$
is zero in this state. The MFM images in figure~\ref{mfmdots}b and
figure~\ref{mfmdots}c confirm the 100 \% magnetic remanence, since
after magnetisation all dots produce a dark signal after
saturation in $H<0$ (figure~\ref{mfmdots}b, $m<0$), or they occur
as bright spots after saturation in $H>0$ (figure~\ref{mfmdots}c,
$m>0$). Drift effects of the MFM tip during the scan cause the
weaker signal of the dots at the bottom of
figure~\ref{mfmdots}b.\\ MFM measurements were also carried out on
the Co/Pt antidot array, shown in figure~\ref{mfmantidots}.
After demagnetisation, a band domain structure in the sample is
clearly visible. The domains are observed as either bright or dark
contrast, which can be associated with $m$ either pointing up or
down. The contrast appearing at the antidots themselves as
white/black objects is possibly due to tip effects because of the
topography. The remanent states are shown in the insets (a) and
(b) of figure~\ref{mfmantidots}, and were obtained after
magnetising the sample in fields of $H=+10$~kOe and $H=-10$~kOe,
respectively. In the inset (a), bright spots are observed at the
position of the antidots which appear due to the mutually opposite
direction of the stray field above the Co/Pt multilayer and above
the antidots. If the Co/Pt multilayer is magnetised in the
opposite direction, see figure~\ref{mfmantidots} inset (b), the
spots have a dark contrast due to the reversed polarity of the
magnetic stray field.

\section{Flux pinning experiments}
In all flux pinning experiments, both the magnetic moments $m$ of
the Co/Pt multilayer and the applied field $H$ are perpendicular
to the sample surface. This leads to interesting magnetic
interactions between the magnetic nanostructures and flux lines,
depending on the mutual orientation of $m$ and $H$.

\subsection{Pinning properties of magnetic dots}
\begin{figure}[tbp]
\centerline{\epsfig{file=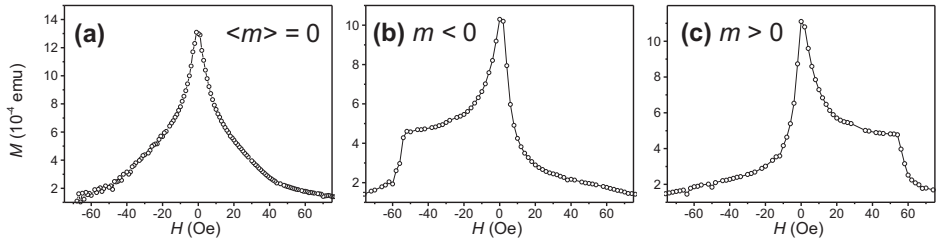}}
\caption{Upper half of the magnetisation loops $M(H)$ measured at
$T=6.61$~K of a 50~nm Pb film on the square array of
Pt(2.5~nm)/[Co(0.4~nm)/Pt(0.9~nm)]$_{10}$ dots after demagnetising
the dots perpendicular to the substrate (a); after magnetising the
dots perpendicular to the substrate in -10~kOe (b) and +10~kOe
(c).} \label{squiddots}
\end{figure}

The pinning properties were studied by SQUID magnetisation
measurements $M(H)$. Figure~\ref{squiddots} shows the upper
branches of the magnetisation curves of the Pb film for $\langle
m\rangle=0$, $m<0$ and $m>0$.
These three magnetic states correspond to the MFM images presented
in figures~\ref{mfmdots}a, b, and c, respectively. During the
measurements, the magnetic state of the Co/Pt dots is preserved
because {\em $H$ is always much smaller than the coercive field
$H_{c}$ of the Co/Pt dots}. The most obvious feature of
figures~\ref{squiddots}b and c is the {\em clear asymmetry of the
M(H) curves with respect to the sign of the applied field}. A
matching effect at the first matching field $H_{1}=\phi_{0}/(600
\textrm{nm})^{2}=57.4$~Oe and an enhancement of $M$ are observed
for aligned $H$ and $m$, whereas no matching effects and a smaller
$M$ can be seen when $H$ and $m$ have opposite polarity. At
$H_{1}$, the field generates exactly one flux quantum $\phi_{0}$
per unit cell of the dot array. The same asymmetry occurs in the
lower branches of the $M(H)$ curves (not shown). The results
presented in figures~\ref{squiddots}b and c indicate that the
pinning force of the dots is much stronger when $m$ and $H$ have
the same polarity.\\ The $M(H)$ curve for the dot array in the
demagnetised state can be seen in figure~\ref{squiddots}a. No
asymmetry and no matching effects appear for this magnetic state.

\subsection{Pinning properties of magnetic antidots}
\begin{figure}[tbp]
\centerline{\epsfig{file=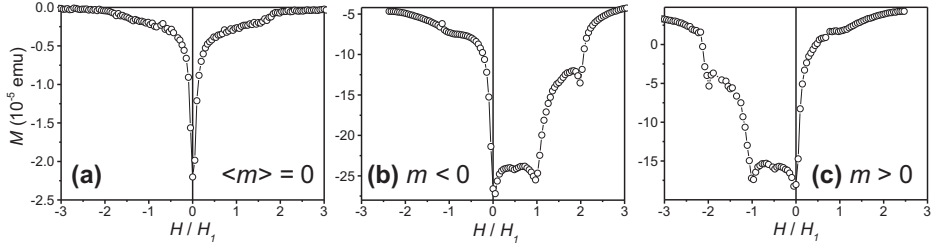}}
\caption{Lower half of the magnetisation loops $M(H/H_{1})$
measured at $T=7.05$~K of a 50~nm Pb film on top of the antidot
array (a) after demagnetisation of the sample perpendicular to the
substrate; (b) and (c) after magnetising the sample perpendicular
to the substrate in -10~kOe and +10~kOe.} \label{squidAD}
\end{figure}

The lower branches of the $M(H)$ curves of the Pb film on top of
the Co/Pt multilayer with a square array of antidots is shown in
figure~\ref{squidAD} for the three different magnetic states
(a)~$\langle m\rangle=0$, (b)~$m<0$ and (c)~$m>0$.
The field axes were normalized to the first matching field
$H_{1}=\phi_{0}/(1 \mu\textrm{m})^{2}=20.67$~Oe. Also for this
sample, a strong asymmetry can be seen in the $M(H)$ curves
presented in figures~\ref{squidAD}b and c. When $H$ and $m$ have
the opposite polarity, a larger $M$ and clear matching effects are
observed at $H/H_{1}=1/2$, $1$ and $2$ for $m<0$ and at
$H/H_{1}=-1/2$, $-1$ and $-2$ for $m>0$, whereas for the same
polarity of $H$ and $m$, a smaller value of $M$ is obtained and
only weak deviations from the smooth curves are visible at
$H/H_{1}=-1$ for $m<0$ and at $H/H_{1}=1$ for $m>0$. In the
$\langle m\rangle=0$ state the $M(H)$ curve has a symmetric shape
with respect to $H$. The magnitude of $M$ is significantly smaller
than in the magnetised states.

\section{Discussion}
\begin{figure}[tbp]
\centerline{\epsfig{file=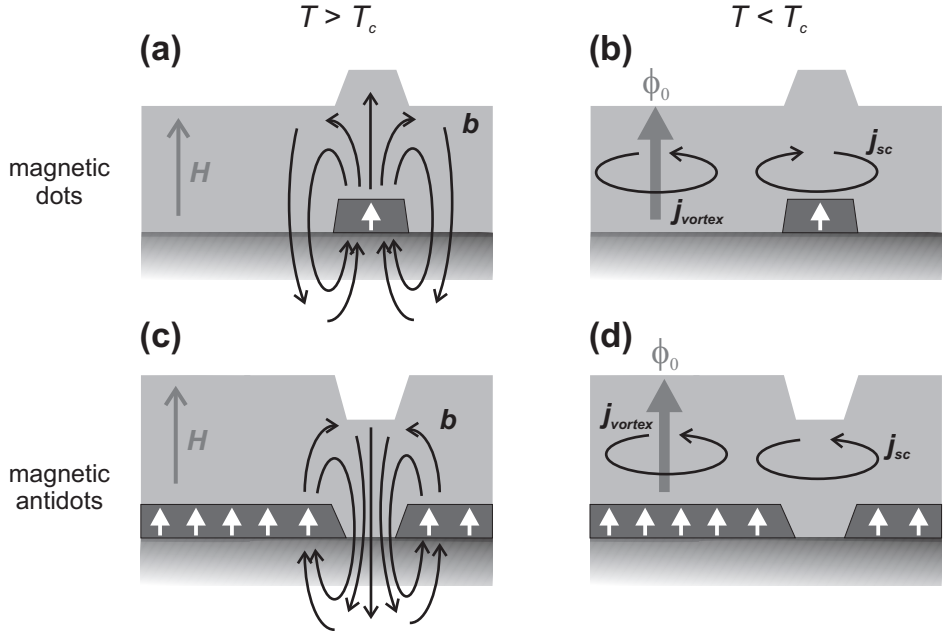}}
\caption{Schematic drawing to illustrate the behaviour of
supercurrents $j_{sc}$ and the stray field $b$ of the magnetic
nanostructures in the $m>0$ state (a)~above $T_{c}$ for the
magnetic dots, (b)~below $T_{c}$ for the magnetic dots (c)~above
$T_{c}$ for the magnetic antidots, and (d)~below $T_{c}$ for the
magnetic antidots. Above $T_{c}$, $b$ can penetrate the Pb film
without inducing supercurrents. Below $T_{c}$, $b$ induces
supercurrents $j_{sc}$, which interact with the supercurrents
$j_{vortex}$ around vortices that are generated by an applied
field.} \label{schema}
\end{figure}

We will now discuss the flux pinning potential that is created in
the superconducting film by the magnetic nanostructures. A lot of
different terms contribute to this potential: non-magnetic
contributions like the corrugated surface of the Pb film as well
as magnetic ones like the high magnetic permeability of the
ferromagnet \cite{martin}, the direction and magnitude of the
magnetic moment \cite{morgan,margriet}, the local stray field of
the ferromagnet \cite{vanbaelPRB}, and supercurrents induced by
the local stray field \cite{vanbaelPRL}. Because of the asymmetry
of the magnetisation curves, the dominating contribution must be a
vector interaction, depending on the mutual orientation of $H$ and
$m$.\\ We will show that the experiments can be consistently
explained by considering the interaction between flux lines and
the supercurrents induced by the stray field of the magnetic
nanostructures in the superconductor. Suppose that the sample is
magnetised in a large positive field, resulting in a perpendicular
component of the stray field that has a positive value above the
dots (see figure~\ref{schema}a).
Because of fluxoid quantization, this stray field can only
penetrate the superconductor in integer multiples of $\phi_{0}$
below $T_{c}$. We assume that the stray field above the dots is
not large enough to induce non-zero fluxoids in the
superconductor. This means that the supercurrents $j_{sc}$ that
are generated by the stray field will try to screen the field from
the interior of the superconductor. As a result $j_{sc}$ will have
a right-handed sense of rotation above the dots (see
figure~\ref{schema}b). In a positive applied field the
supercurrents around the vortices $j_{vortex}$ have the opposite
sense of rotation as $j_{sc}$. Consequently the vortices are
attracted to the dots, resulting in the pronounced matching effect
when $m$ and $H$ are aligned. On the other hand, flux lines that
are generated by a negative applied field will be repelled from
the dots because $j_{vortex}$ has the same sense of rotation as
$j_{sc}$. The flux lines will occupy the interstitial positions
between the dots where they are weaker pinned. This causes the
absence of matching effects when $H$ and $m$ have opposite
polarity.\\ In the demagnetised state, the dots with $m>0$ and
$m<0$ are randomly distributed over the square array. As a result,
the pinning potential landscape is not periodic anymore and the
pinning force is the same for negative and positive $H$. The lack
of periodicity of the pinning potential is reflected by the
symmetric shape of the $M(H)$ curve shown in
figure~\ref{squiddots}(a).\\ The stray field of the magnetic
antidots in the $m>0$ state has opposite polarity compared to the
magnetic dots in the $m>0$ state (compare figures~\ref{schema}a
and c). Therefore, assuming that also the magnetic antidots do not
induce any fluxoids in the superconductor, $j_{sc}$ above the
antidots have opposite sense of rotation as $j_{sc}$ above the
dots, compare figure~\ref{schema}b with figure~\ref{schema}d.
Consequently the magnetic antidots have opposite flux pinning
properties as the magnetic dots and pronounced matching effects
appear when $H$ and $m$ have opposite polarity. In the $\langle m
\rangle =0$ state, magnetic domains with $m$ either pointing up or
down are present in the sample (see figure~\ref{mfmantidots}).
From the MFM image one can see that these domains are randomly
distributed. This means that the stray field of the sample does
not reflect the periodicity of the antidot array as in the $m>0$
and the $m<0$ state, resulting in the absence of matching effects
and the symmetric $M(H)$ curve shown in figure~\ref{squidAD}a.

\section{Conclusions}
We have studied the pinning properties of a type-II
superconducting film on top of two different types of {\em
magnetic} artificial pinning centres. Such kind of pinning centres
provides a strong pinning potential for the flux lines, yielding
pronounced asymmetric $M(H)$ magnetisation curves for
perpendicularly magnetised samples. This opens the opportunity to
tune the properties of a superconductor by switching between the
magnetic states of the ferromagnetic nanostructures.

\section*{Acknowledgements}
The authors thank R. Jonckheere (IMEC vzw), K. Temst and G.
G\"untherodt for help with sample preparation, D. Buntinx for MFM
measurements and J. Swerts for MOKE measurements. This work is
supported by the Belgian Inter-University Attraction Poles (IUAP)
and Flemish Concerted Research Actions (GOA) programs, by the ESF
"VORTEX" program and by the Fund for Scientific Research-Flanders
(FWO). MJVB is a post-doctoral research fellow of the FWO.

\end{document}